\documentclass[epj]{svjour}

\usepackage{latexsym}
\usepackage{graphicx}
\usepackage{fancyhdr}
\usepackage{amssymb,amsmath,array}

\def\makeheadbox{{%  remove EPJC header
 }}
\def\mytitle{Running of $\alpha_s$ in the MSSM with three-loop accuracy } 
\def\myauthors{Luminita Mihaila}  
\def\mytype{Parallel talk}
\def\mysession{Colliders - Susy Phenomenology}
\def\mytitle{Running of $\alpha_s$ in the MSSM with three-loop accuracy} %Put your title here!
\def\myauthors{Luminita Mihaila}    %Put your name here!
\def\mytype{Contributed Talk}    
\def\mysession{Colliders - SUSY Phenomenology}
\begin{document}
\title{Running of $\alpha_s$ in the MSSM with three-loop accuracy}
\author{Luminita Mihaila
%\inst{1}
%\thanks{\emph{Email:} Insert  Email  of corresponding author here}%
% \and
% Robert Harlander\inst{2} \and Matthias Steinhauser\inst{1}  
}                     % Do not remove
\institute{Institut f{\"u}r Theoretische Teilchenphysik - Universit{\"a}t Karlsruhe\\ 
76128 Karlsruhe - Germany}
%\and Fachbereich C Theoretische Physik - Universit{\"a}t Wuppertal \\
%   42097 Wuppertal - Germany}
\date{}
\abstract{
The evolution of the strong coupling constant $\alpha_s$ from
$M_Z$ to the GUT scale is presented, involving three-loop running and
two-loop decoupling. Accordingly, the two-loop transition from 
the $\overline{\rm MS}$ to the $\overline{\rm DR}$ scheme is
properly taken into account.   
We find that the three-loop effects are comparable to the experimental
uncertainty  for $\alpha_s$.
 % end of PACS codes
} %end of abstract
\maketitle
%DO NOT REMOVE THIS LINE
%

\section{Introduction}
\label{intro}
Supersymmetry (SUSY) is currently believed to play an important role in
physics beyond the Standard Model\\(SM). There are several reasons that
point out the Minimal Supersymmetric Standard Model(MSSM) as a preferred 
 theory describing new physics. The foremost ones are its milder
 divergency structure that solves the naturalness problem and the
 possibility to explain the electroweak supersymmetry breaking as a
 consequence of  radiative corrections. Another compelling argument in
 favour of SUSY is the particle content of the
 MSSM, that leads in a natural way to the unification of the three gauge
 couplings at a high energy scale $\mu\simeq
10^{16}$~GeV, in agreement with  Grand Unification Theories (GUT).
 This observation together with the   consistent predictions made for SM
parameters, such as the top quark mass and the ratio of the bottom
quark to the tau lepton masses, using constraints on the Yukawa sector
of SUSY-GUT models, brought SUSY in the center of the phenomenological studies.  
 
Nevertheless, SUSY can only be an approximate symmetry in
nature and  several scenarios for the mechanism of SUSY breaking
 have been proposed. A possibility to
constrain the type and scale of SUSY breaking is to study, with very
high precision, the relations between the MSSM parameters evaluated at
the electroweak and the GUT scales.  The extrapolation over many orders
of magnitude requires high-precision experimental data at the low energy
scale. A first set of precision measurements is expected from the CERN
Large Hadron Collider (LHC) with an accuracy at the percent
level. A comprehensive high-precision analysis can be performed at the
International Linear Collider (ILC), for which the
estimated experimental accuracy is at the per mill level. In this
respect, it is
necessary that the same precision  is reached
also on the theory side in order to match with the
data~\cite{Aguilar-Saavedra:2005pw}. Running analyses based on full
two-loop renormalization group equations
(RGEs)~\cite{Martin:1993yx,Jack:1994kd} for
all parameters and one-loop threshold corrections~\cite{Pierce:1996zz}
are currently implemented in the public programs
ISAJET~\cite{Paige:2003mg}, SOFTSUSY~\cite{Allanach:2001kg},
SPHENO~\cite{Porod:2003um}, SuSpect~\cite{Djouadi:2002ze}. The agreement
between the different codes is in general within one
percent~\cite{Allanach:2003jw}. A first three-loop running analysis,
based, however, only on one-loop threshold effects, was carried out in
Ref.~\cite{Ferreira:1996ug}.

In this talk, we report on 
the evaluation of the
strong coupling $\alpha_s$  in  MSSM, based on 
three-loop RGEs\\~\cite{Jack:1996vg} and two-loop threshold
corrections~\cite{Harlander:2005wm}. 
On the one hand, the  three-loop corrections
reduce  significantly the dependence on the scale at which heavy
particles are integrated out~\cite{Harlander:2007}. On the other
hand, they are 
essential for  phenomenological studies, because they are as large as, or
greater than, the effects induced by the current experimental accuracy of
$\alpha_s(M_Z)$~\cite{Bethke:2006ac}.
Additionally, we compare the
predictions obtained within the above 
mentioned approach with those based on the
leading-logarithmic (LL) approximation suggested in
Ref.~\cite{Aguilar-Saavedra:2005pw}.

\section{Evaluation of $\alpha_s(\mu_{\rm GUT})$ from
  $\alpha_s(M_Z)$ }
\label{sec:1} 

The aim of this study is to compute  $\alpha_s$  at a high-energy scale
$\mu\simeq{\cal O}(\mu_{\rm GUT})$, starting from the strong coupling
constant at the mass of the 
$Z$ boson $M_Z$. We denote this parameter
 $\alpha_s^{\overline{\rm MS},(5)}(M_Z) $  to specify that the
underlying theory is QCD  with five active
flavours and  $\overline{\rm MS}$ is the renormalization scheme.  
The value of $\alpha_s(\mu_{\rm GUT})$ is the
strong coupling constant in the MSSM renormalized in the
$\overline{\rm DR}$-scheme, that we denote as $\alpha_s^{\overline{\rm DR},(\rm
    full)}(\mu_{\rm GUT})$.
 For the evaluation of $\alpha_s^{\overline{\rm
    DR},(\rm full)}$ from  $\alpha_s^{\overline{\rm MS},(n_f)}$ we
   follow the ``common scale approach''~\cite{Baer:2005pv}, which
  requires  a unique scale for  the matching between QCD and MSSM. More precisely,
 for mass independent renormalization schemes like 
$\overline{\rm MS}$ or $\overline{\rm DR}$,
the decoupling of heavy particles has to be performed explicitely.  In
practice, this means that intermediate effective theories are introduced
by integrating out the heavy degrees of freedom. One may separately
integrate out every particle at its individual threshold (``step
approximation''), a method suited for SUSY models with a severely split
mass spectrum. But the intermediate effective theories with ``smaller''
symmetry raise the problem of introducing new couplings, each governed
by its own RGE. 
To overcome this difficulty, for SUSY models with roughly degenerate mass
spectrum at the scale $\tilde M$, one can consider the MSSM as the full
theory that 
is valid from the GUT scale $\mu_{\rm GUT}$ down to $\tilde{M}$, which we
assume to be around $1$\,TeV. 

For the running analysis of the strong coupling constant,
we can distinguish four individual steps  that we detail below.
%\begin{equation}
%  \begin{split}
%    \alpha_s^{\overline{\rm MS},(n_f)}(M_Z)\quad
%    &\stackrel{(i)}{\to}\quad \alpha_s^{\overline{\rm
%    MS},(n_f)}(\mu_{\rm dec})\quad 
%    \stackrel{(ii)}{\to}\quad \alpha_s^{\overline{\rm
%    DR},(n_f)}(\mu_{\rm dec})\\ 
%    &\stackrel{(iii)}{\to} \quad\alpha_s^{\overline{\rm DR},(\rm
%    full)}(\mu_{\rm dec})\quad 
%    \stackrel{(iv)}{\to}\quad \alpha_s^{\overline{\rm DR},(\rm
%    full)}(\mu_{\rm GUT})\,. 
%  \end{split}
%  \label{eq::asrundec}
%\end{equation}
%The individual steps require: $(i)$ $\beta(\alpha_s)$ in QCD through
%three loops, $(ii)$ the $\overline{\rm MS}$--$\overline{\rm DR}$
%relation through order $\alpha_s^2$, $(iii)$ decoupling of the SUSY
%particles through order
%$\alpha_s^2$, and $(iv)$  $\beta(\alpha_s)$ through three loops in
%SUSY-QCD.
\begin{enumerate}
\item  Running of  $\alpha_s^{\overline{\rm
    MS},(n_f)}$ from $\mu=M_Z$ to $\mu=\mu_{\rm dec}$. \\
The energy dependence of the strong coupling constant 
is governed by the RGEs
\begin{eqnarray}
\mu^2\frac{\rm d}{\rm d \mu^2} \alpha_s(\mu^2) = \beta(\alpha_s)\,, \nonumber\\
\beta(\alpha_s) = -\alpha_s^2\sum_{n\geq 0}\beta_n\alpha_s^n\,.
\label{eq::betadef}
\end{eqnarray}
In QCD with $n_f$ quark flavours, the $\beta$
function is known through four
loops  both in
the $\overline{\rm MS}$~\cite{vanRitbergen:1997va,Czakon:2004bu} and
the  $\overline{\rm DR}$-scheme~\cite{Harlander:2006xq}.
\item  Conversion of $\alpha_s^{\overline{\rm
    MS},(n_f)}(\mu_{\rm dec})$ to $\alpha_s^{\overline{\rm
    DR},(n_f)}(\mu_{\rm dec})$.\\
For the three-loop running analysis we are focusing on, one needs to
evaluate the dependence of  $\alpha_s$  values in the $\overline{\rm
  DR}$ scheme from those evaluated in  $\overline{\rm MS}$ scheme with
    two loops accuracy~\cite{Harlander:2006xq} 
\begin{eqnarray}
\!\! \!\! \alpha_s^{\overline{\rm    MS}} = \alpha_s^{\overline{\rm
    DR}}\left[1-\frac{\alpha_s^{\overline{\rm DR}}}{4\pi} 
    -  5\frac{ ( \alpha_s^{\overline{\rm DR}})^2}{4 \pi^2}
    + n_f\frac{ \alpha_s^{\overline{\rm DR}}\alpha_e^{}}{12\pi^2}
   \right]
  \label{eq::asMS2DR}
\end{eqnarray}
Here, the following notations have been used $\alpha_s^{\overline{\rm DR}} \equiv
    \alpha_s^{\overline{\rm DR},(n_f)}(\mu)$ and 
$\alpha_s^{\overline{\rm    MS}} \equiv \alpha_s^{\overline{\rm
    MS},(n_f)}(\mu)$. $\alpha_e{} \equiv \alpha_e^{(n_f)}(\mu)$ is
one of the so-called evanescent coupling constants that occur when
$\overline{\rm DR}$ is applied to non supersymmetric theories (QCD in
this case). In particular, it describes the coupling of the
$2\varepsilon$-dimensional components (so-called $\varepsilon$-scalars)
of the gluon to a quark. It is an unphysical parameter that must
decouple from any prediction for physical observables. We also used this property 
as a consistency check for our method. As mentioned above, we assume that QCD is obtained by
integrating out the heavy degrees of freedom (squarks and gluinos) from
SUSY-QCD. In this case, the evanescent couplings are uniquely determined
    by the matching conditions between the two theories, that we discuss below.

\item Matching of $\alpha_s^{\overline{\rm
    DR},(n_f)}(\mu_{\rm dec})$ and $\alpha_s^{\overline{\rm DR},(\rm
    full)}(\mu_{\rm dec})$.\\
Integrating out all SUSY particles at the common scale of SUSY mass spectrum, 
one directly obtains the SM as the effective theory, valid at low
    energies.
The transition between the two theories can be done at an
arbitrary decoupling scale $\mu$:
\begin{eqnarray}
\alpha_s^{\overline{\rm DR},(n_f)}(\mu) &=&
 \zeta_s^{(n_f)}\,\alpha_s^{\overline{\rm DR},(\rm full)}(\mu)\nonumber\\
 \alpha_e^{(n_f)}(\mu)&=& \zeta_e^{(n_f)} \alpha_e^{(\rm full)}(\mu)\,.
\label{eq::asdec}
\end{eqnarray}
$\zeta_s$ and $\zeta_e$  depend logarithmically on the scale $\mu$,
which is why one generally chooses $\mu\sim \tilde M$.  
In Eq.~(\ref{eq::asdec}), $n_f=6$ means
that only the SUSY particles are integrated out, while for $n_f=5$ at
the same time the top quark is integrated out. \\
In a
supersymmetric theory,  SUSY requires that $ \alpha_e^{(\rm
  full)}(\mu)=\alpha_s^{\overline{\rm DR},(\rm full)}(\mu)$ at any scale. 
Let us remark that $\alpha_s^{\overline{\rm  DR},(full)}(\mu_{\rm dec})$
 is not known {\it a priori} and one cannot use
Eq.~(\ref{eq::asdec}) directly in order to derive
$\alpha_e^{(5)}$. Rather, we start with a trial value for 
 $\alpha_s^{\overline{\rm  DR},(full)}(\mu_{\rm dec})$
 and obtain the corresponding $\alpha_e^{(5)}(\mu_{\rm dec})$ as well as 
$\alpha_s^{\overline{\rm DR},(5)}(\mu_{\rm dec})$
through Eq.~(\ref{eq::asdec}). 
 Then we evaluate $\alpha_s^{\overline{\rm MS},(5)}(\mu_{\rm dec})$ through
Eq.~(\ref{eq::asMS2DR}), and from that $\alpha_s^{\overline{\rm MS},(5)}(M_Z)$.  The trial value
for $\alpha_s^{\overline{\rm  DR},(full)}(\mu_{\rm dec})$ is systematically varied until the
resulting $\alpha_s^{\overline{\rm MS},(5)}(M_Z)$ agrees with the
experimental input.

\item Running of $\alpha_s^{\overline{\rm DR},(\rm full)}$ from
  $\mu=\mu_{\rm dec}$ to $\mu=\mu_{\rm GUT}$.\\
The energy dependence of the strong coupling constant 
is in this case governed by the MSSM RGEs. In SUSY-QCD, the
$\beta$ function has been evaluated in the $\overline{\rm DR}$-scheme 
through three loops~\cite{Jack:1996vg}.
\end{enumerate}

Assembling the above mentioned steps, we can predict the value of $\alpha_s(\mu_{\rm
  GUT})$ with up to three-loop accuracy.
This procedure is implemented in
most of the present codes computing the SUSY 
spectrum~\cite{Porod:2003um,Allanach:2001kg,Djouadi:2002ze}  by applying
the one-loop approximation of Eq.~(\ref{eq::asdec}) and 
setting $n_f=5$ and $\mu=M_Z$.
The advantage of this procedure as compared to a multi-scale approach is
that the RGEs are only one-dimensional  and that for $\alpha_e$ one can apply
Eq.~(\ref{eq::asdec}).

Let us note that in principle it is possible to decouple the top quark
separately. The only new ingredient needed
is the decoupling constant for going from five to six quark flavours in
the $\overline{\rm MS}$ scheme. In any case, for a mass spectrum
as given by the benchmark point
SPS1a$^\prime$~\cite{Aguilar-Saavedra:2005pw}, for example, the separate
decoupling of the top quark implies a numerically small effect. This can
also be established by comparing ``Scenario D'' and ``Scenario C'' in
Ref.~\cite{Harlander:2005wm}.

The phenomenological significance of the three-loop order corrections is
discussed in detail in the next section.

\section{ Numerical results}
\label{sec:2} 
\begin{figure}
\includegraphics[width=0.49\textwidth,angle=0]{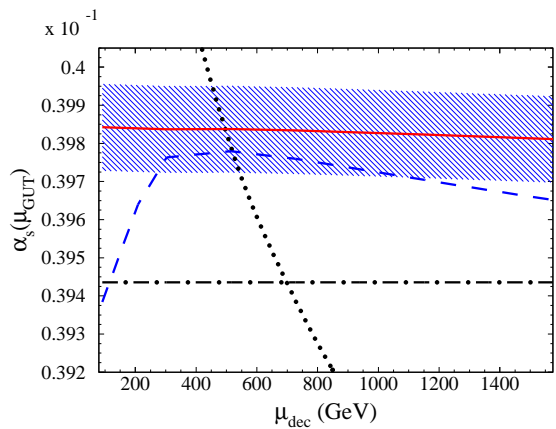}
\caption{$\alpha_s(\mu_{\rm GUT})$ as a function of $\mu_{\rm dec}$.}
\label{fig:asgut}       % Give a unique label
\end{figure}

The result for 
$\alpha_s^{\overline{\rm DR}}(\mu_{\rm GUT}=10^{16}~\mbox{GeV})$,
obtained using $M_Z = 91.1876~\mbox{GeV}$,
$ m_t=170.9\pm 1.9~\mbox{GeV}\,,$
$\alpha_s^{\overline{\rm MS}}(M_Z) = 0.1189\,,$ and
$\tilde{M} = m_{\tilde{q}} = m_{\tilde{g}} = 1000\,\mbox{GeV}$
as input parameters is shown if Figure~\ref{fig:asgut}.  The dotted,
dashed and solid line are based on the approach described above, where
$n$-loop running is combined with $(n-1)$-loop decoupling, as it is
required for consistency ($n=1,2,3$, respectively). 
We find a nice convergence when going from one to three
loops, with a very weakly $\mu_{\rm dec}$--dependent result at
three-loop order. 
For comparison, we show the result (the dash-dotted line) obtained from
the formula given in 
Eq.~(21) of Ref~\cite{Aguilar-Saavedra:2005pw}. It corresponds to the
resummed one-loop contributions originating from both the
change of scheme  and the decoupling of heavy particles. However, the
difference between our three-loop result with two-loop decoupling (upper
solid line) and the one-loop formula given in
Ref.\,\cite{Aguilar-Saavedra:2005pw} exceeds the experimental
uncertainty by almost a factor of four for sensible values of
$\mu_{\rm dec}$.  This uncertainty is indicated by the
hatched band, derived from $\delta\alpha_s(M_Z)= \pm
0.001$~\cite{Bethke:2006ac}. The formulae of
Ref.\,\cite{Aguilar-Saavedra:2005pw} should therefore be taken 
only as rough estimates.

\begin{figure}
\includegraphics[width=0.49\textwidth,angle=0]{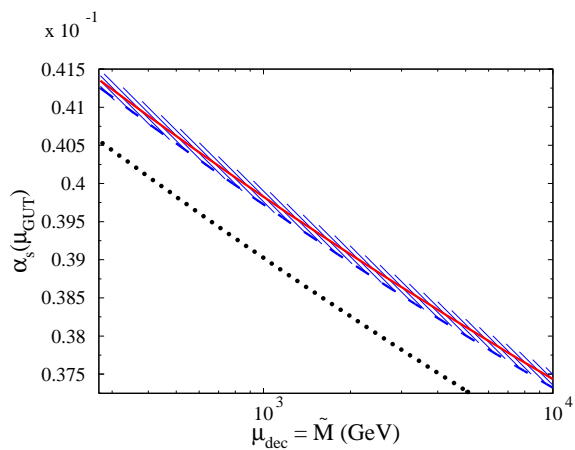}
\caption{$\alpha_s(\mu_{\rm GUT})$ as a function of $\tilde{M}$.}
\label{fig:asgutmsusy}       
\end{figure}

In Figure~\ref{fig:asgutmsusy} we show $\alpha_s(\mu_{\rm GUT})$
as a function of $\tilde{M}$ where $\mu_{\rm dec}=\tilde{M}$ has been
adopted. Dotted, dashed and full curve correspond again to the one-,
two- and three-loop analysis and the uncertainty form $\alpha_s(M_Z)$
is indicated by the hatched band. One observes a variation of
10\% as $\tilde{M}$ is varied between 
$100$~GeV and $10$~TeV. This shows that the actual SUSY scale can
significantly influence the unification, respectively, the
non-unification behaviour of the three couplings 
at the GUT scale. 

\section{Conclusions}

We have used recent three- and two-loop results for the $\beta$
functions  and the decoupling
coefficients, respectively, in order to derive $\alpha_s^{\overline{\rm DR}}(\mu_{\rm
  GUT})$ from $\alpha_s^{\overline{\rm MS}}(M_Z)$  at three-loop level.

It turns out that the three-loop terms are numerically significant. 
 The dependence on where the SUSY spectrum is decoupled becomes
 particularly flat in this case. 
The theoretical uncertainty is expected to be negligible w.r.t.\ the
uncertainty induced by the experimental input values. 

Comparing our results and methods to the literature, we find that the
issue of evanescent couplings has either been ignored (by assuming
$\alpha_e=\alpha_s$) or circumvented by decoupling the SUSY spectrum at
$\mu_{\rm dec}=M_Z$. We find that at one- and two-loop level, this choice does
not allow for a good approximation of the higher order effects, if one
assumes the SUSY partner masses to be of the order of 1\,TeV.

In consequence, we
 recommend that  phenomenological studies concerning the
 implications of low energy data on Grand Unification
should be done at three-loop level.

\bigskip
\noindent
{\large\bf Acknowledgements}\\ 
The author thanks R.~Harlander and M.~Steinhauser for a fruitful
collaboration on this topic.
This work was supported by the DFG through SFB/TR~9.

%
% BibTeX users please use
% \bibliographystyle{}
% \bibliography{}

\begin{thebibliography}{999}


%
% mihaila_ref.tex -- generated by sortref-2.3.6  
% ((C) R. Harlander, http://www.robert-harlander.de/software/)
% on Wed Sep 12 11:11:09 CEST 2007
%
\bibitem{Aguilar-Saavedra:2005pw}
  J.~A.~Aguilar-Saavedra {\it et al.},
  Eur.\ Phys.\ J.\ C {\bf 46} (2006) 43.

%2
\bibitem{Martin:1993yx}
  S.~P.~Martin and M.~T.~Vaughn,
  Phys.\ Lett.\ B {\bf 318} (1993) 331. 

%3
\bibitem{Jack:1994kd}
  I.~Jack and D.~R.~T.~Jones,
  Phys.\ Lett.\  B {\bf 333} (1994) 372.

%4
\bibitem{Pierce:1996zz}
  D.~M.~Pierce, J.~A.~Bagger, K.~T.~Matchev and R.~J.~Zhang,
  Nucl.\ Phys.\ B {\bf 491} (1997) 3. 

%5
\bibitem{Paige:2003mg}
  F.~E.~Paige, S.~D.~Protopopescu, H.~Baer and X.~Tata,
  [arXiv:hep-ph/0312045].

%6
\bibitem{Allanach:2001kg}
  B.~C.~Allanach,
  Comput.\ Phys.\ Commun.\  {\bf 143} (2002) 305.

%7
\bibitem{Porod:2003um}
  W.~Porod,
  Comput.\ Phys.\ Commun.\  {\bf 153} (2003) 275.

%8
\bibitem{Djouadi:2002ze}
  A.~Djouadi, J.~L.~Kneur and G.~Moultaka,
  Comput.\ Phys.\ Commun.\  {\bf 176} (2007) 426.

%9
\bibitem{Allanach:2003jw}
  B.~C.~Allanach, S.~Kraml and W.~Porod,
  JHEP {\bf 0303} (2003) 016.

%10
\bibitem{Ferreira:1996ug}
  P.~M.~Ferreira, I.~Jack and D.~R.~T.~Jones,
  Phys.\ Lett.\  B {\bf 387} (1996) 80.

%11
\bibitem{Jack:1996vg}
  I.~Jack, D.~R.~T.~Jones and C.~G.~North,
  Phys.\ Lett.\  B {\bf 386} (1996) 138.

%12
\bibitem{Harlander:2005wm}
  R.~Harlander, L.~Mihaila and M.~Steinhauser,
  Phys.\ Rev.\ D {\bf 72} (2005) 095009.

%13
\bibitem{Harlander:2007}
  R.~V.~Harlander, L.~Mihaila and M.~Steinhauser,
  Phys.\ Rev.\  D {\bf 76} (2007) 055002.

%14
\bibitem{Bethke:2006ac}
  S.~Bethke,
  Prog.\ Part.\ Nucl.\ Phys.\  {\bf 58} (2007) 351.
%
%18
\bibitem{Baer:2005pv}
  H.~Baer, J.~Ferrandis, S.~Kraml and W.~Porod,
  Phys.\ Rev.\  D {\bf 73} (2006) 015010.

%

%15
\bibitem{vanRitbergen:1997va}
  T.~van Ritbergen, J.~A.~M.~Vermaseren and S.~A.~Larin,
  Phys.\ Lett.\  B {\bf 400} (1997) 379.

%16
\bibitem{Czakon:2004bu}
  M.~Czakon,
  Nucl.\ Phys.\  B {\bf 710} (2005) 485.

%17
\bibitem{Harlander:2006xq}
  R.~V.~Harlander, D.~R.~T.~Jones, P.~Kant, L.~Mihaila and M.~Steinhauser,
  JHEP {\bf 0612} (2006) 024.
%
%



\end{thebibliography}

\end{document}